# An empirical estimate of the electricity supply curve from market outcomes

Jorge Sánchez Canales [a*], Alice Lixuan Xu [a], Chiara Fusar Bassini [ab], Lynn H. Kaack [ab], Lion Hirth [ac]

[a] *Centre for Sustainability, Hertie School, Friedrichstraße 180, 10117 Berlin, Germany*
[a] *Data Science Lab, Hertie School, Friedrichstraße 180, 10117 Berlin, Germany*
[a] *Neon Neue Energieökonomik GmbH, Karl-Marx-Platz 12, 12043 Berlin, Germany*
[*] *Corresponding author:* j.sanchez-canales@hertie-school.org

**Abstract**     Researchers and electricity sector practitioners frequently require the supply curve of electricity markets and the price elasticity of supply for purposes such as price forecasting, policy analyses or market power assessment. It is common practice to construct supply curves from engineering data such as installed capacity and fuel prices. In this study, we propose a data-driven methodology to estimate the supply curve of electricity market empirically, i.e. from observed prices and quantities without further modeling assumptions. Due to the massive swings in fuel prices during the European energy crisis, a central task is detecting periods of stable supply curves. To this end, we implement two alternative clustering methods, one based on the fundamental drivers of electricity supply and the other directly on observed market outcomes. We apply our methods to the German electricity market between 2019 and 2024. We find that both approaches identify almost identical regimes shifts, supporting the idea of stable supply regimes stemming from stable drivers. Supply conditions are often stable for extended periods, but evolved rapidly during the energy crisis, triggering a rapid succession of regimes. Fuel prices were the dominant drivers of regime shifts, while conventional plant availability and the nuclear phase-out play a comparatively minor role. Our approach produces empirical supply curves suitable for causal inference and counterfactual analysis of market outcomes.

# 1. Introduction

Supply curves are one of the best-known and widest-used tools of microeconomics, featured in virtually all introductory Economics classes. They describe how much of a commodity producers are willing to supply for a given price. With the demand curve, they establish the market equilibrium. With fully identified supply and demand curves, one can generate potential market outcomes (equilibrium quantity and price) and thus conduct counterfactual causal inference, as discussed by Tiedemann et al. (2024) for the context of electricity markets. Such procedures allow to estimate the welfare effects following an intervention and are thus the foundation of cost-benefit and ex-ante policy evaluation (Mishan & Quah, 2020; Yang et al., 2014). More recently, elasticities have been applied to assess the marginal value of public funds in energy and climate policy (Hahn et al., 2024; Hendren, 2016). Estimating counterfactual prices is also central to detecting conditions for market power and to measure the incentives for market power abuse, a topic particularly important given the nature of electricity markets (Xu et al., 2025).

In electricity markets, supply curves play a central role across policy, industry, and academic applications, from assessing market power and market reform impacts to informing investment and dispatch decisions. However, existing studies that determine supply curves are often either based on engineering data such as information about conversion efficiency and installed capacity, or on hourly bid data. These approaches face important limitations when it comes to causal inference: either the amount of data or modeling assumptions, the hour-specific nature of biding curves or the problems of omitted variable bias and functional form of statistical models limit the potential of these curves for counterfactual analysis. In order to overcome these shortcomings, our goal in this study is to estimate *empirical inverse supply curves* that express the hourly price of wholesale electricity in the day-ahead market as a function of equilibrium system quantity in a way that is useful for counterfactual analysis of market outcomes in general, similar to the way in which many studies estimate demand curves empirically.

Instead of relying on an instrumental variable approach, as is common in the demand-side studies, we employ a different identification strategy that exploits the structure of electricity markets. Wind and solar supply, as well as demand, fluctuate vastly on short time scales. For instance, residual demand (system load minus renewable energy) can fluctuate by up to 60 GW within a single day. In contrast, the conventional supply curve is stable over short horizons, since it is mainly influenced by slowly evolving factors such as fuel prices, hydro storage, and thermal plant availability. We argue that, as long as conventional supply is stable and after accounting for variable renewable generation, changes in the observed equilibrium prices and quantities must be exclusively the result of high-frequency shifts in demand. If there are enough observations and such a period of stability persists for a long enough time and there is sufficient demand-side variation, then we can recover the underlying supply relationship without requiring external instruments just by observing these equilibrium points over time (Tiedemann, 2025).

However, we also know that supply conditions can shift substantially over longer periods. Recent examples of this is the year 2020, when Covid-19 depressed fuel prices, and 2021 to 2023, when reduced Russian gas exports sharply increased them. We hence aim to identify periods with a constant thermal supply curve, which we denote *regimes*, and points in time where it changes, which we refer to as *regime shifts*.

We develop two alternative methods to empirically detect these regime shifts. Both methods employ time-series segmentation algorithms built on the pruned exact linear time (PELT) segmentation algorithm by Killick et al. (2012) but approach the problem from opposite and complementary directions. The first, "cause-driven" approach, seeks to hold the fundamental drivers of supply constant *before* estimating supply curves from the data. The second, "effect-driven" approach, identifies regularities in market outcomes that imply the existence of stable supply regimes that generated them.

We use the case of Germany between 2019 and 2024 as an application of our approach. We find that our cause and effect driven methods identify similar dates as regime shifts. We argue that this 1) supports that both regimes and regime shifts are meaningful concepts to use in empirical research of electricity supply; 2) shows that fuel prices were the main drivers behind regime shifts in our sample period; and 3) supports the concrete regime shifts that we identify within our sample period. Finally, the estimated supply curves within each regime are generally a good fit for the data, especially for those regimes identified via the effect-driven approach. This suggests that the estimated supply curves are correctly identified.

## 2. Related literature

Researchers and practitioners often make use of electricity supply curves, explicitly or implicitly. In empirical studies, there are three main ways in which supply is operationalized: via numerical models based on engineering data and commodity prices, via observed bid curves in spot market auctions, and via supply-side variables (i.e., renewable generation, fuel prices, etc.) used in statistical models. We point out the gap in the literature by explaining how these existing approaches fall short of producing an empirical supply curve to use for counterfactual analysis.

First, numerical models simulate market outcomes by solving optimization problems (e.g. cost minimization or welfare maximization) subject to techno-economic constraints (DeCarolis et al., 2017; Pfenninger et al., 2014; Schill and Zerrahn, 2018). They are primarily used to assess forward-looking scenarios[1] and in some cases forecasting market outcomes[2]. Typically, these models do not estimate the supply relationships from empirical data but rather impose them through their techno-economic assumptions. However, the so-called inverse optimization models are designed to do just that: starting with a fundamental, bottom-up model of the electricity market, they learn its parameters from observed market outcomes instead of getting them as exogenous inputs. These learnt relationships could be considered empirical supply curves.

---

[1] Such as decarbonization targets (Müsgens, 2020), bidding zone splits (Czock, 2025; Zinke, 2023), the integration of new technologies like electrolyzers, batteries, electric vehicles, and heat pumps (Bernath et al., 2021; Ruhnau, 2022; Ruhnau et al., 2020), the effects of carbon pricing (Ruhnau et al., 2022) or renewable expansion (Hirth, 2013).
[2] For example, fundamental models based on approximated supply curves can be used to forecast day-ahead prices (Kulakov and Ziel, 2019; Ziel and Steinert, 2016), particularly for long-term estimates (Ghelasi and Ziel, 2025a).

Important examples are the works by Ruiz et al. (2013) and Mitridati and Pinson (2018), though crucially, they only apply their methods to simulated data, highlighting the computational intensity and limited empirical applicability of their methods; and the works by Sahraei-Ardakani et al. (2015) and Ghelasi and Ziel (2025b), who use real data in applications to the US and Germany, respectively. All of these studies empirically estimate supply curves by inferring the key parameters in their respective models from the observed market outcomes. However, this approach remains heavily dependent on a pre-defined model structure, susceptible to misspecifications in both modelled and unmodelled interactions. Ultimately, the supply curve is not directly observed but is the output of an optimization problem calibrated on real-world data.

The second strand of literature corresponds to articles that build supply curves from observed bids in the day-ahead auction. There are many examples in this literature. For instance, one relevant body of work uses bid curves to analyze market power (Ciarreta Antuñano and Espinosa, 2010; Reguant, 2014; Wolak, 2003; Wolfram, 1998). Supply curves derived from bids have also been used for other applications, such as price forecasting (Li et al., 2024). In empirical studies, Fabra and Reguant (2014) used them to assess the pass-through of carbon prices to consumers, and Robinson et al. (2023) used them to estimate the impacts on prices of the Iberian exception mechanism during the European energy crisis.

While these studies provide valuable insights, using observed bids is not the same as estimating what we refer to as the aggregate supply curve. Fundamentally, this approach conflates the day-ahead auction –a specific market mechanism on a single trading platform– with the broader market's supply.[3] Ultimately, this approach leads to having different supply curves for each trading period, curves which reflect the willingness to sell in a particular moment and are heavily influenced by short-term factors like unit commitment and strategic positioning. This results in a highly context-dependent relationship that may not represent the stable, medium-run supply curve necessary for credible counterfactual analysis.

Lastly, there is a third strand of statistical and econometric models that use supply drivers to fit or forecast market outcomes. The most relevant subset of this literature are studies that aim at identifying causal effects or structural coefficients directly pertaining to the supply side of the market. For instance, Fabra and Reguant (2014) made a model to estimate the share of the pass-through of carbon prices to consumers; Jahns et al. (2020) regress prices on hydro reservoir production and other factors to estimate the aggregate supply curve of hydro reservoirs in Norway; and Tselika et al. (2024) quantify the impact of short-term changes in renewable energy production on electricity prices. While these models can be extremely useful in isolating the effect of certain drivers of supply on equilibrium prices, these are not supply functions. Moreover, this general approach is susceptible to functional-form misspecification (i.e., how the other variables affect the core relationship between prices and quantities), as the estimated curve is dependent on the chosen parametric form and potentially affected by omitted variable bias.

---

[3] This can be a reasonable assumption in some markets. For example, in Spain, where both of these studies take place, there is a single trading platform, OMIE. The volume traded in the day-ahead auction organized by OMIE corresponds to around 70% of the physical volume on the grid. In some other countries, where there are several trading platforms, this assumption becomes more problematic.

Methodologically, we are most aligned with the study by Bataille et al. (2019). Their goal is to estimate the counterfactual price impact of capacity withholding. After considering and dismissing using fuel prices as controls, they study the bilateral relationship between residual load and market equilibrium prices. However, their analysis is fundamentally static, implicitly assuming a single, stable supply regime throughout their study period (the year 2016 in the Austro-German market). This limits its applicability to periods without any major structural shocks. Crucially, they claim that despite the endogeneity of market outcomes, they do not need an instrumental variable because of the short-run inelasticity of demand. We further show that to retrieve supply curves one does not need to make that assumption (see Section 3.1. below, particularly Figure 1).

We can learn from all these studies. First, inverse optimization problems and the work by Bataille et al. (2019) show that one can estimate the supply curve from empirical data. Second, the statistical literature has produced many insights into which are empirically the most relevant drivers of electricity supply. However, there is a lack of a methodology that is purely empirical and can estimate dynamic supply curves that can be used for counterfactual analysis without assuming out endogeneity by claiming that demand is inelastic. Numerical models are rigid and data-intensive. Even when calibrated with real-world data, they are not truly reduced-form and remain vulnerable to misspecification. Bid-based approaches mix day-ahead auction behavior with broader market dynamics, producing curves that reflect short-term bidding strategies rather than a stable supply relationship, and often require proprietary granular data. Econometric models typically impose parametric structures and many controls, creating risks of misspecification or omitted variables. They are useful for studying mechanisms but often model price formation rather than the core $P = f(Q)$ relationship. The few studies that estimate this relationship directly, such as Bataille et al. (2019), are static and cannot account for regime shifts.

Our study addresses this gap by introducing a dynamic reduced-form approach. Similar to Bataille et al. (2019), we argue that the aggregate supply curve is a stable empirical relationship that can be directly observed from the market equilibrium, but only within a well-defined *supply regime*. To do so, we first provide a method for estimating the reduced-form supply curve as a flexible, piecewise-linear function of residual load. This approach minimizes a priori assumptions, requires only publicly available market outcome data, and yields an intuitive P = f(Q) relationship that is easily applicable for policy counterfactuals. Second, we solve the problem of static relationships by developing a time-series segmentation framework that empirically identifies shifts in supply regimes. Thus, our analysis remains robust through periods of market shock. In summary, we complement existing literature by offering a practical, reduced-form tool for causal inference and counterfactual analysis that is both empirical and adaptable to different and changing markets.

## 3. Method

### *3.1. Identification strategy*

Our goal in this study is to recover the *empirical inverse supply curve* from observed market equilibriums (pairs of prices and quantities). The core challenge in estimating a structural supply relationship from market data is endogeneity: observed prices and quantities are simultaneously determined by supply and demand. While the research in empirical supply estimates is lacking (in favor of the conventional bottom-up engineering approaches), there is ample research in the analogous problem of demand estimation. In that literature, the established solution to endogeneity is to use instrumental variables, such as wind power generation, to isolate exogenous price variation (Bönte et al., 2015; Fabra et al., 2021; Hirth et al., 2024; Knaut and Paulus, 2016; Tiedemann et al., 2024).

However, in this article we identify the supply curve without an instrument. This is not because we ignore the problem of endogeneity, but because our identification strategy rests on a different identifying assumption, derived from the dynamics of electricity markets. We operate under the notion that within a stable supply regime, where the thermal supply curve remains constant, the entire variation in equilibrium price-quantity pairs is driven by exogenous shifts in demand. This is consistent with the condition under which Tiedemann (2025) argues that an instrument may be unnecessary: when the exogenous variation in the independent variable is large relative to the confounding variation from the structural error term.

Formally, we want to estimate the inverse supply curve of the type $P_t = f(Q_t, \epsilon_t)$, where $t$ refers to each hour.

- The independent variable $Q_t$ refers simultaneously to the quantity demanded, the quantity supplied and the (observed) equilibrium quantity. Its exogenous variation comes from the multitude of factors that shift hourly demand (weather, time-of-day, holidays, randomness) independently of supply-side factors.
- The structural error term $\epsilon_t$ represents unobserved hourly supply shocks (e.g., unforeseen outages, changing trade conditions).

It is clear that the supply curve that maps this relationship, $f(\cdot)$, will depend on several factors, such as fuel prices and the available generation capacity. The core identification argument is that if we successfully isolate a period where the function remains stable, the exogenous demand-driven variation in quantity will be the dominant source of fluctuation, making the potential bias from any residual supply-side endogeneity (stemming from a correlation between $Q_t$ and $\epsilon_t$) negligible. Instead of conditioning on the relevant drivers, we identify these periods of stable supply by segmenting the time series. Consequently, the set of observed equilibrium points closely reflects the underlying supply curve, allowing us to recover it directly by fitting a curve (Figure 1).

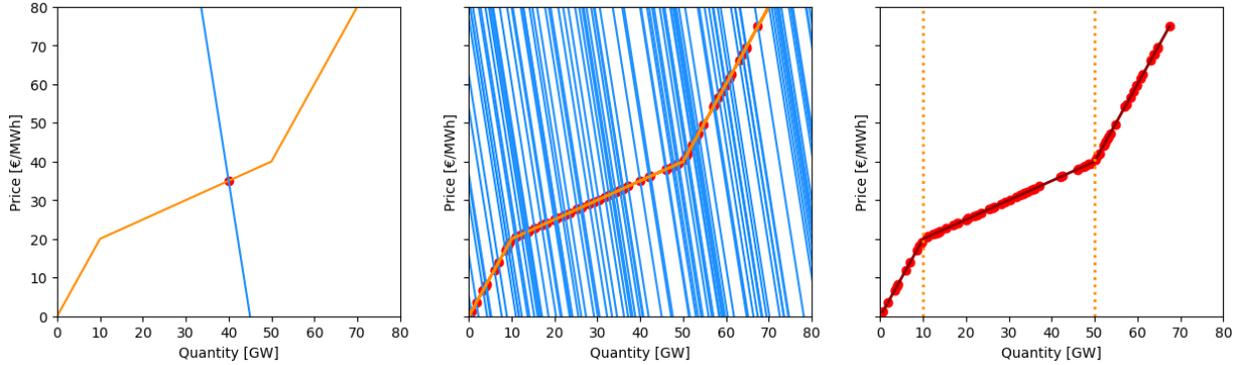

**Figure 1.** Identification strategy as illustrated with synthetic data. **Left:** A demand curve (blue) and a supply curve (orange) determine an equilibrium price–quantity pair (red dot). **Center:** Several realizations of the demand curve generate multiple equilibria, reflecting changes in demand intercepts (e.g., hourly and daily variations) while assuming constant price responsiveness of demand. **Right:** From these observed equilibria, a piecewise-linear function can be fitted to recover the underlying supply curve.

This operation of segmenting the time-series is analogous to "conditioning" on certain variables in a regression setting, as done by the econometric models discussed in Section 2.1. This way, we reduce the problem of functional form misspecification, since conditioning via time-series segmentation is non-parametric. This approach is not a shortcut to avoid using an instrument, but a deliberate methodological choice justified by the specific and well-understood economics of electricity supply. Moreover, it has the advantage of allowing us to fit the whole range of the supply curve, and not only the parts where an instrument could provide coverage.

We fit a supply curve for each period of stable supply, modeling price as a function of residual load, following the approach in Bataille et al. (2019). Whereas Bataille et al. (2019) use a cubic specification, we employ a data-driven piecewise linear fit to minimize functional assumptions about the shape of the supply curve. This approach remains consistent with the fundamental representation of supply in electricity economics as a piecewise or step function reflecting the merit-order of marginal costs (Stoft, 2002). Piecewise supply functions have been applied both in empirical estimations of supply curves from market outcomes (Jahns et al., 2020; Sahraei-Ardakani et al., 2015) and in calibrated bottom-up models (Ghelasi and Ziel, 2025b). Our implementation is a piecewise linear fitting algorithm, adapted from Jekel and Venter (2019), which estimates breakpoint locations and segments slopes that minimize squared errors while ensuring the curve remains non-negative (weakly monotonic), thus capturing the economic intuition that higher quantities should not lead to lower prices.

We use the term *regimes* to denote the periods of stable supply and we identify significant changes in behavior of the time series in different time intervals as *regime shifts*, or *changepoints*. The concept of regimes in electricity price formation is not new, but its empirical use has been limited for two main reasons. First, given the slow-moving nature of fundamental drivers such as fuel prices, available capacity, and cross-border interconnection, it was long reason-able to assume a single supply regime was responsible for price formation. This was the case until large-scale shocks such as the European energy crisis of 2022 made this assumption unjustifiable. Second, most existing studies that examine regime shifts use pre-post identification, working under the assumption that the timing of the changepoint is known ex-ante, since often it is tied to an identifiable market event such as a policy reform or a planned plant shutdown.[4] In contrast, we identify regime shifts ex-post directly from the data without making further assumptions.

We model regimes as time-continuous, instead of treating every hour independently, for several reasons. First, it follows the tradition of the previous pre-post studies. After the decommissioning of a power plant, we cannot go back to the previous regime. Second, it conceptually follows the intuition from sequential updating in modern time series analysis (Box and Jenkins, 1970): each regime builds on updated priors from the previous one, reflecting an evolving rather than reversible market process. In other words, we do not assume that markets can revert to an identical past state after a shock, even if similar conditions re-emerge.[5] Third, restricting regimes to contiguous time segments substantially reduces the complexity of the clustering problem. Allowing the grouping together of non-adjacent days would require considering all possible partitions of the time series, a search space that grows exponentially with the number of observations. By contrast, enforcing temporal continuity reduces the problem to segmenting a sequence, which can be solved efficiently using dynamic programming. For these three reasons we identify regimes as continuous segments of the time series and examine the issue of seasonality or reversion to previous regimes through a comparison of regime similarities in Section 4.

*3.2. Time series segmentation*

Since we want to learn regimes, we must use a changepoint detection algorithm. While approximate changepoint detection algorithms, such as binary segmentation (Scott and Knott, 1974; Sen and Srivastava, 1975), and its variants are often adopted due to their low computational cost, they do not guarantee an exact solution. On the other hand, exact methods such as neighborhood segmentation (Auger and Lawrence, 1989) and optimal partitioning (Jackson et al., 2005) have a quadratic computational cost. Our method builds on the pruned exact linear time (PELT) segmentation algorithm (Killick et al., 2012), which combines exact solution retrieval with a linear[6] runtime, making it suitable for the computation of regimes over a long sample period (2192 days).

PELT attempts to find a set of $m$ changepoints $\tau_1, \tau_2, \ldots, \tau_m$ that minimizes the *total cost function*:

---

[4] For instance, nuclear power plant shutdown in California (Davis and Hausman, 2016) and in Germany (Jarvis et al., 2022), the introduction of flow-based market coupling in Central-western Europe (Ovaere et al., 2023), or the Iberian exception (Robinson et al., 2023).

[5] Consider the Iberian exception as an example. One could argue that the market could in principle have reverted to the same regime that existed before the crisis once the intervention ended. However, such policy interventions could well alter the beliefs of market actors, for instance about the regulatory responses in face of scarcity pricing. Even if all observable conditions later resemble those of an earlier period, would the bidding behavior be the same?

[6] This assumes that evaluating the cost function is O(1), i.e., takes a constant amount of time. However, the actual runtime of PELT depends on the chosen cost function, which in our case may require more computation.

$$C = \sum_{i=0}^{m} C_m \left(y_{(\tau_i+1):\tau_{i+1}}\right) + \beta m$$

defined as the sum of *segment cost functions* $C_m$ plus a penalty term, which is the linear product of the number of segments $m$ and a user-defined penalty term $\beta$. PELT uses a dynamic programming recursion, with a pruning trick, which discards candidate changepoint that cannot be part of an optimal segmentation. The operational steps can be found as pseudocode in Appendix A under Algorithm 1.

For identifying the regimes, we use two different approaches, denoted as *cause-driven* and *effect-driven*. First, in the cause-driven approach, we seek to segment the time-series of the drivers of supply. This approach stems from the assumption that in periods where the fundamental drivers of supply are stable themselves, the supply curve must be as well. The second one, the effect-driven approach, works from the opposite direction: under the assumption that stable supply curves must produce stable relations of price-quantity in equilibrium, we look for the stable curve in market outcomes by fitting supply curves. Since clustering in general is an underspecified problem, we employ both approaches to validate our results. Operationally, they share the same overarching algorithm, PELT, and the differ in terms of the input data[7] and the specific *segment cost function* $C_m(\cdot)$ which they use. These details are explained in their respective sections below. Figure 2 summarizes the conceptual structure of our method of identifying supply functions from market outcomes.

### 3.2.1. Cause-driven

The cause-driven approach identifies supply regimes by jointly segmenting the time series of the fundamental drivers that determine the supply curve's position and shape. In this section we explain the segment cost function and the input data that we use in our cause-driven approach to time series segmentation. We focus on the three drivers for which we can construct relevant, high-frequency, market-wide time series: variable renewable generation, fuel and emission costs, and available capacity.

Variable renewable energy (wind, solar and hydro energy without storage) is not used as an input for the clustering algorithm, but it plays a crucial role in defining our independent variable: to estimate the supply curve we use residual load, that is, total system load minus variable renewable generation. This choice is motivated by the merit order effect (Sensfuß et al., 2008). Due to their near-zero marginal cost of production and, in many cases, subsidies, renewable generators displace conventional generators, thereby lowering market-clearing prices. Put differently, since renewables are typically dispatched first, they shift the supply curve to the right. There is ample evidence of this effect (Cludius et al., 2014; Figueiredo and Silva, 2019; Hirth, 2013; Ketterer, 2014; Kolb et al., 2020; Martin De Lagarde and Lantz, 2018). Therefore, the main reason behind using residual load is that it "corrects" the merit-order effect of renewable energy, effectively "un-shifting" the supply curve to recover the underlying relationship between price and the demand faced by dispatchable generators. This is the standard procedure in many studies (e.g., Bataille et al., 2019). For our specific identification strategy, it has the added benefit that it removes a major source of high-frequency supply-side variation (fluctuating renewable infeed) that would otherwise confound our estimate and it increases the exogenous, demand-driven variation used for identification, as residual load exhibits greater hourly and daily volatility than total load alone.

---

[7] All data sources are listed in Section 3.3.

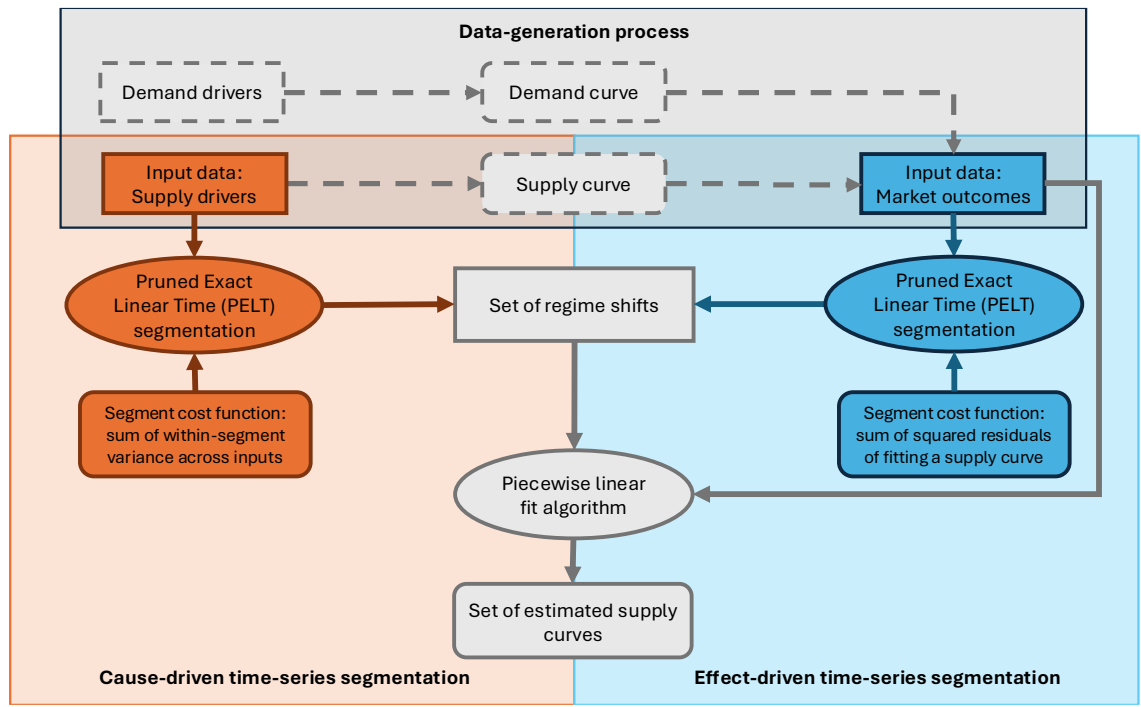

**Figure 2.** Conceptual figure of our method to identify supply functions from observed market outcomes. Dashed lines denote unobserved processes. Solid lines denote our method. **Top:** Underlying market-outcome generation process, where supply and demand drivers determine the corresponding curves. The intersection of the curves yields equilibrium prices and quantities. **Left:** cause-driven time-series segmentation, which identifies supply regimes based on observed supply-side drivers using a PELT algorithm with a variance-based cost function. **Right:** effect-driven segmentation, which instead infers supply regimes directly from market outcomes via a PELT algorithm using a piecewise linear fit cost function. Both approaches yield regime shifts that are used to estimate one supply curve per identified regime.

We use carbon-adjusted fuel prices as input for our time segmentation algorithm. Fuel and carbon prices are among the most important determinants of electricity supply and price formation in wholesale markets, as they are the main components of the marginal cost of thermal generation. Empirical research shows that the prices of crude oil, natural gas, and coal are the primary fuel-related drivers of wholesale electricity prices, with natural gas being the main marginal price-setter in most European markets in recent years (Ding et al., 2020; Furió and Chuliá, 2012; Mohammadi, 2009; Zakeri et al., 2023). For Germany's capacity mix, natural gas and hard coal are the primary marginal price-setters. Regarding carbon prices, a substantial body of work finds an almost complete pass-through from EU Allowance (EUA) prices to wholesale electricity prices (Bai and Okullo, 2023; Bunn and Fezzi, 2007; Fabra and Reguant, 2014). We therefore construct daily, carbon-adjusted fuel price indices in EUR/MWh thermal for coal and natural gas. This incorporates the cost of the fuel itself and the cost of the corresponding EUA, calculated using standard carbon intensity factors.

We also use thermal available capacity as an input for our time segmentation algorithm. Electricity supply ultimately depends on the physical capacity of power plants. To be part of the supply curve, generation units must both exist and be operationally available. Capacity thus reflects not only long-term investment and retirement decisions (Fleten et al., 2019; Mills et al., 2017), but also short-term availability. When part of the installed capacity is unavailable, the market must rely on more expensive plants to meet demand, leading to higher prices for the same quantity supplied (Bergler et al., 2017; Durmaz et al., 2024). Therefore, we use a daily time series of available conventional capacity, as reported by power plant operators, as the second input for our cause-driven approach (Fusar Bassini et al., 2025). This variable captures commissioning and de-commissioning of individual power plants, but also outages due to scheduled maintenance, forced unavailability, and other operational constraints that physically limit the supply offered to the wholesale electricity market, potentially including even market power abuse. We use available capacity divided by fuel and technology, as well as aggregated, as shown in Table 1 below.

Our approach does not explicitly account for some other potentially relevant supply-side drivers, notably trade flows and temporal dynamics. Although empirical evidence suggests that these factors can influence short-run supply,[8] we do not include them as inputs for three main reasons. First, their marginal impact is likely small relative to the dominant drivers (fuel prices, available thermal capacity and renewable generation). Second, incorporating temporal dynamics would require a different functional form for the supply curve, complicating identification. Third, trade patterns have high hourly volatility (mostly in response to weather and demand patterns), making them ill-suited as predictors of medium-run supply regimes. We therefore treat these omitted factors as short-run deviations that do not structurally characterize supply regimes.

As discussed in Section 3.1, successful identification of the supply curve from market outcomes requires abundant demand-side variation and close to no remaining supply-side variation. Implicitly, our method relies on the assumption that supply regimes are exclusively determined by fuel costs and available capacity and that the influence of variable renewable generation can be removed by using residual load, while the influence of all other drivers averages out to a small, normally distributed error term. We state this assumption explicitly so that readers can judge whether the remaining influence of the unobserved drivers is plausibly negligible or whether their omission threatens the validity of our cause-based regime identification.

---

[8] See for instance (Böckers and Heimeshoff, 2014; Bunn and Gianfreda, 2010; Gugler et al., 2018; Keles et al., 2020; Newbery et al., 2016; Ovaere et al., 2023) for some studies on trade and Kumar and Saini (2022) or Tselika et al (2024) for the existence of temporal dynamics in supply.

An advantage of using two complementary approaches is that the potential impact of these omissions is empirically testable. If developments in unobserved drivers cause significant, persistent regime shifts, our complementary effect-driven approach (Section 3.2.2.) should be able to detect them. If, on the other hand, the regimes identified by both methods agree, then that provides supporting evidence that our assumptions for the cause-driven approach are reasonable for the purpose of identifying medium-run supply stability.

Regarding input data, we run several specifications, which we denote with the prefix $C$ to signal they are cause-driven (Table 1). Specification C1 uses only the carbon-adjusted prices of coal and gas, since fuels are likely the most relevant drivers for the period we study. Specification C2 includes total available capacity, to represent changes to the overall stock. Specification C3 is more refined: we include separately the available capacity of nuclear and lignite. The reason is that these two are baseload technologies: they have lower marginal costs, so they are located at the beginning of the merit order. If some of these units are unavailable, the whole supply curve would be shifted to the left. Moreover, in our sample period, Germany decommissioned all nuclear power plants as part of a planned phase-out. Because of the very different scales (price and capacity), C2 and C3 use the inputs as normalized series, whereas C1 uses prices in €/MWh.

|    | Prices | | Available capacity | | | Series scale | |
|----|--------|---|--------------------|---|---|--------------|---|
|    | Coal | Gas | Total | Nuclear | Lignite | Levels | Normalized |
| C1 | X | X |   |   |   | X |   |
| C2 | X | X | X |   |   |   | X |
| C3 | X | X |   | X | X |   | X |

**Table 1.** Input data and variable scale used for each cause-driven specification.

The segment cost function is shared across all cause-driven specifications. Let each potential segment $s_m = (\tau_{j-1} + 1, \ldots, \tau_j)$ include all observations between changepoints $\tau_{j-1}$ and $\tau_j$. Let $x_{i,t}$ denote the value of driver $i$ (e.g., carbon-adjusted fuel price, available capacity) at time $t$, and $I$ the number of drivers included. Then, for the cause-driven approach, the segment cost function can be defined as the sum of the within-segment variance (or equivalently, the sum of squared deviations from the segment mean) across all series:

$$C_m = \sum_{i=1}^{I} \sum_{t=\tau_{j-1}+1}^{\tau_j} (x_{i,t} - \bar{x}_{i,m})^2, \quad \text{with } \bar{x}_{i,m} = \frac{1}{\tau_j - \tau_{j-1}} \sum_{t=\tau_{j-1}+1}^{\tau_j} x_{i,t}.$$

### 3.2.2. Effect-driven

The effect-driven approach characterizes regimes from the other side of the data generating process: market outcomes. Here, we propose that observed periods of stability in market outcomes must correspond to a single supply regime.

As visually motivated in Figure 1, short-term variations of demand levels can be used to recover the full shape of the supply curve from market equilibria observed in the day-ahead market, i.e., pairs of clearing prices and residual load. The important condition is that supply remains stable for the observed period: attempting to fit a single supply curve to equilibria generated by different supply regimes means estimating the wrong structural relationship; this misspecification would typically result in a poor fit. Thus, we exploit the idea that a stable supply curve should produce a narrow set of market outcomes by using a measure of goodness of fit, the sum of squared residuals, as the *segment cost function* of a proposed regime. In other words, for $p_t$ the observed price, $q_t$ the (residual) load and $\hat{f}_m(\cdot)$ the piecewise linear supply curve estimated for segment *m*, the segment cost function is the sum of squared errors from that fit:

$$C_m = \sum_{t=\tau_{j-1}+1}^{\tau_j} \left(p_t - \hat{f}_m(q_t)\right)^2$$

We build on the piecewise linear fit algorithm (PWLF), which –given a number of breakpoints *k*– finds the slopes and breakpoint locations that minimizes the sum of squared errors of the model (Jekel and Venter, 2019). Our implementation enforces the additional condition that the slopes of the linear functions should be non-negative, as (partially) negative supply curves would imply that increases quantity would result in decreasing prices. Details of the implementation are illustrated in Algorithm 2 in Appendix A.

We run several effect-driven specifications, denoted with the prefix *E*. In this case, they differ not in the input data but in how flexible the fitted supply curve is. E1 fits a single straight line (no breakpoints). E2 fits a piecewise linear curve with exactly two breakpoints. E3 allows up to two breakpoints but chooses them adaptively: it adds breakpoints one at a time and keeps the new one only if it reduces the sum of squared residuals by at least 1%. Otherwise, it stops and uses the simpler curve.

### 3.2.3. Stopping criterion

The PELT algorithm that we adapt for both of our approaches requires a penalty associated with adding new regimes to avoid overfitting. A common approach would be to use information criteria, such as BIC or AIC, but in the case of this algorithm, for the pruning step to work properly, the penalty must be linear with respect to the number of regimes (Killick et al., 2012). This linear penalty is referred to as $\beta$ in Algorithm 1.

While running all our specifications, we modify $\beta$ to find the minimum number of regimes that is necessary to explain a certain threshold of variance that would be left unexplained by the baseline model. We calculate this baseline error by applying the segment cost function to the overall period, meaning that for each specification, the baseline unexplained variation is different. For example, for specification C1, it is the sum of the variance of the overall time series of coal and natural gas prices, whereas for E1, it is the sum of squared residuals of fitting all market equilibria in a single linear regression. Figure 3 shows the level of unexplained variance as a function of the number of regimes used for specification C1, as an example. For each specification and each threshold of explained residuals, we identify a different number of required regimes, presented in Table 2.

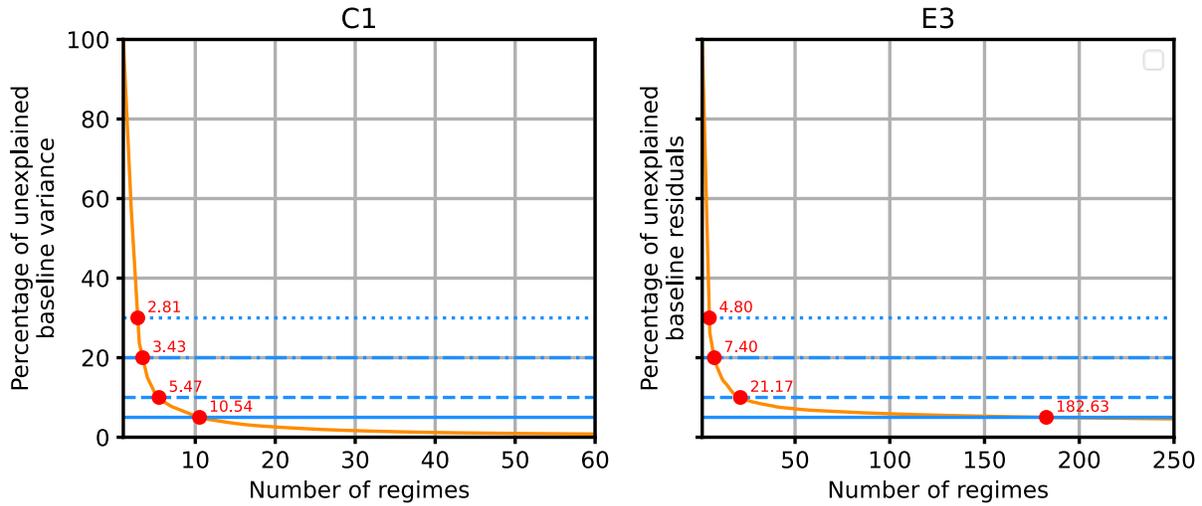

**Figure 3**. Example of variance ratio of unexplained residuals across different numbers of regimes for specifications C1 and E3. The orange line shows the share of unexplained variance (sum of squared residuals under each number of regimes divided by the baseline case of one regime) as the number of regimes increases. Horizontal lines indicate benchmark thresholds for residual variance (5%, 10%, 20%, and 30%). Markers denote the intersection points between the variance curve and these thresholds, identifying the regime counts at which unexplained variance falls below each benchmark.

| SPECIFICATION | 70% | 80% | 90% | 95% |
|---|---|---|---|---|
| **C1** | 3 | 4 | 6 | 11 |
| **C2** | 3 | 4 | 9 | 20 |
| **C3** | 5 | 9 | 28 | 105 |
| **E1** | 5 | 9 | 27 | 631* (673) |
| **E2** | 5 | 8 | 22 | 175* (177) |
| **E3** | 5 | 8 | 23 | 184* (193) |

**Table 2**. Number of regimes required to explain different levels of variance in our six-year sample (2019-2024). Numbers with * come from linear interpolation, with the closest observed number of clusters in between brackets.

We see from Table 3 that explaining residual variation in effect-driven specifications is more difficult. This is to be expected, given the respective cost functions and the amount of input data. Whereas cause-driven uses daily data (and therefore has 2192 points), effect-driven segments daily but fits curves on hourly data (with a total of 52,608 market equilibrium observations). We chose to use 11 regimes (or 10 regime shifts) as the comparison point. This corresponds to 95% of explained variance in C1, that is, in the fuel prices series. We include the results of alternative number of regimes (5 and 20) in Appendix B (Figures 10 and 11).

*3.3. Data*

Our sample includes 6 years of data, from January 1st, 2019, to December 31st, 2024. We obtain hourly wholesale electricity market data (price, load, renewable generation) from ENTSO-E Transparency Platform (ENTSO-E, 2024) and hourly generation unit availability data from EEX Transparency Platform (EEX, 2024; Fusar Bassini, 2025). The fuel price indices we use in this study have daily resolution and include carbon emissions in the EU ETS (International Carbon Action Partnership, 2025), natural gas at Dutch TTF futures (Investing.com, 2025a), and hard coal Argus-McCloskey futures (Investing.com, 2025b). Figures 8 and 9 in Appendix B visualize the data input data for cause-driven specifications.

## 4. Results

In this section, we present our main results. We begin by comparing the regime shifts identified by each specification. We then zoom in on the gas crisis to show how relevant fuel prices are in shaping these shifts. After that, we show examples of the resulting supply slopes, which illustrate how the different specifications interpret the underlying supply curve. Finally, we return to our assumption of time-continuous regimes and test whether the data supports it.

We run all our specifications with 11 regimes. Figure 4 shows a timeline for each one, with markers indicating the identified regime shifts. We find that there is a strong alignment between the shifts identified by C1 (which relies only on carbon-adjusted fuel prices in levels) and all effect-driven specifications. By contrast, C2 and C3, the cause-driven specifications that include available capacity (and inputs are therefore normalized), show noticeably different patterns. Still, some changepoints (for instance, the one at the beginning of 2023) appear across all specifications.

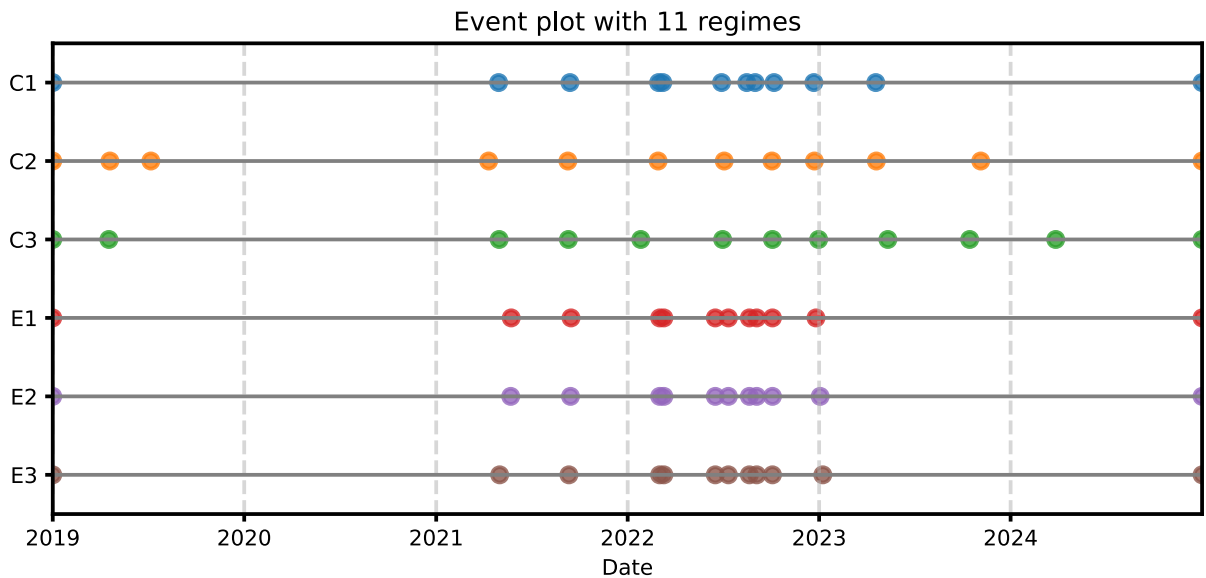

**Figure 4**. Event plot of the whole sample (2019-2024) showing the point in time where each specification identified the regime shifts when looking for 11 regimes.

There are several things we can learn from this exercise. First, the strong alignment between C1 and all effect-driven specifications suggests that fuel prices are the main drivers of regime changes. This is consistent with the idea that supply regimes remain stable as long as underlying cost drivers remain stable. Second, although seasonality matters for available capacity and is accordingly captured by C2 and C3 (see Figure 11 in Appendix B) it does not seem to translate into meaningful changes in the shape of the empirical supply curve. This is an interesting result on its own: it suggests that generators schedule maintenance in a way that avoids major distortions in the supply regime. Third, the high degree of agreement across all effect-driven methods indicates that the functional-form flexibility matters surprisingly little. Whether the supply curve within a regime is estimated as linear (E1) or piecewise linear with multiple segments (E2 and E3), the timing of regime shifts barely changes. This suggests that most of the variation in the period we study reflects movements in the level or steepness of the supply curve rather than changes in its internal shape.

Because most of the regime shifts in C1 and in the effect-driven specifications occur during the gas crisis, we zoom in on this period. Figure 5 shows carbon-adjusted natural gas prices alongside the regime shifts between 2021 and 2023. While the shifts do not line up perfectly day-by-day – and instead typically offset by only a few days– they follow the same major movements. This pattern supports the idea that gas and coal prices were the key drivers. The small timing differences could reflect delays in price passthrough, differences between the international index and effective natural gas prices in Germany, or other short-run operational frictions.

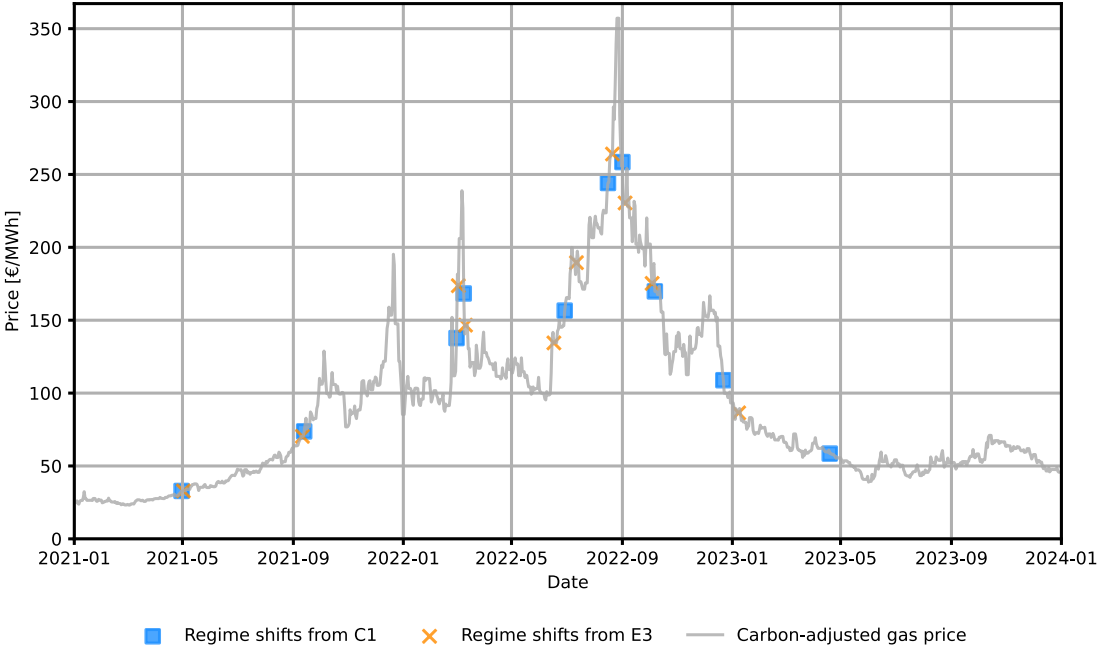

**Figure 5**. Event plot, zooming in the gas crisis (2021-2023). Grey line depicts a daily index of carbon-adjusted natural gas prices. Blue squares show regime shifts identified by C1 (cause-driven specification using only fuel prices as input). Orange crosses show regime shifts identified by E3 (effect-driven specification using a flexible curve fit).

Identifying regime shifts is an intermediate step toward our main objective: recovering empirical supply curves that can be used for counterfactual price analysis. Figure 6 shows the piecewise-linear supply curves obtained using the regimes from E3. These curves illustrate how much the level and the shape of the supply curve changed across regimes.

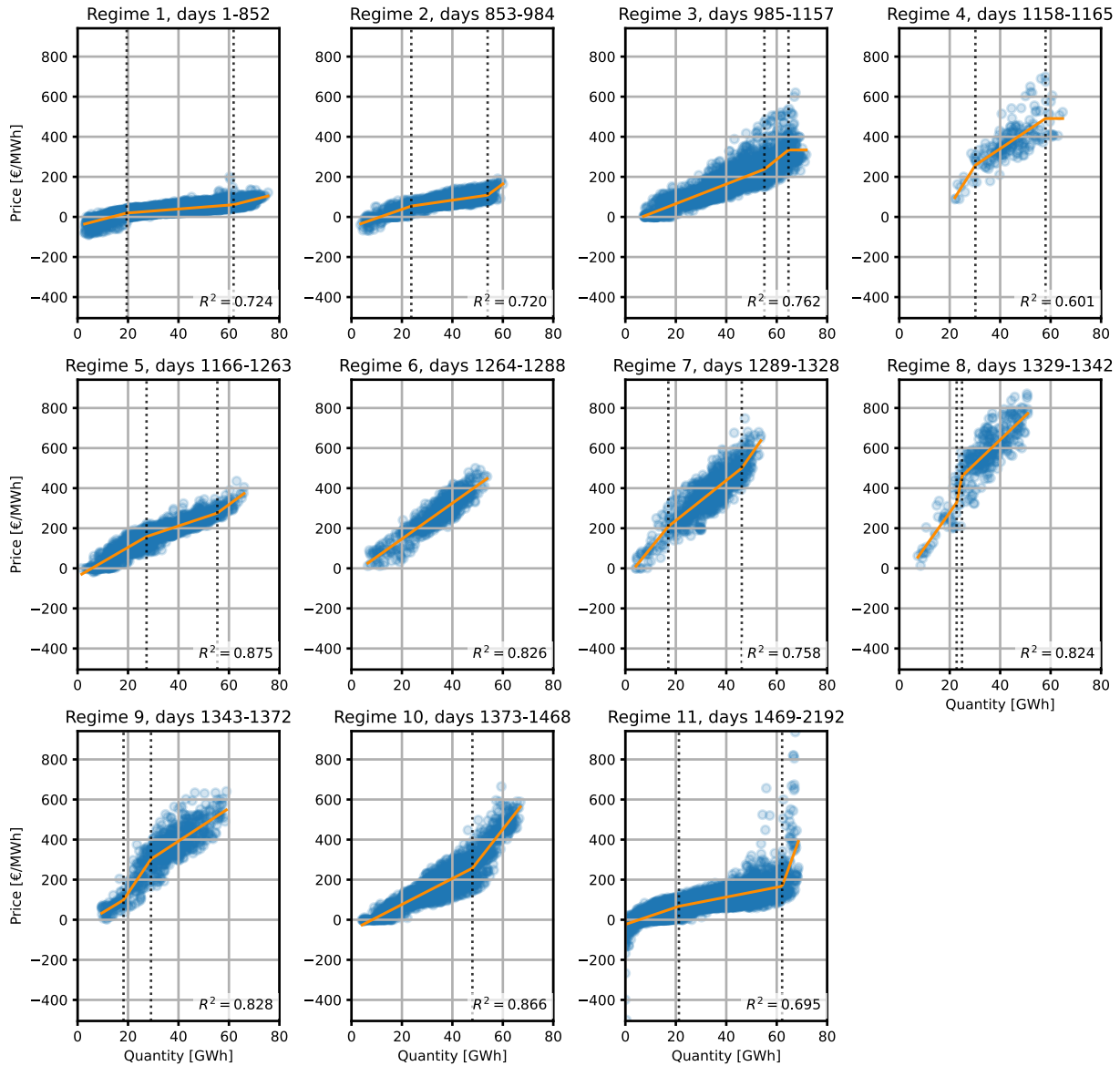

**Figure 6**. Piecewise linear supply curves resulting from using the regimes identified by E3. Blue dots represent each hourly equilibrium of the day-ahead market auction. In orange, the estimated piecewise linear supply curve. Dotted vertical lines mark the breakpoints of the piecewise linear functions.

Figure 6 confirms the intuition that dividing the observed market equilibria into regimes is meaningful. The method clearly picks up structure: the sets of market equilibria form tight clusters, and the fitted curves track them well (in general, the $R^2$ is above 0.7). This tightness is crucial for us to identify supply without endogeneity from demand (see Section 3.1). It can be argued that there is some remaining supply-side variation, but most of it has been accounted for while leaving a substantial demand-side variation, necessary to "map out" the supply curve: across most plots we see sufficient support along quantities (x-axis), whereas looking at prices, clear relationships emerge. On the contrary, another immediate observation is the large difference in regime sizes. Regimes 1 and 11 cover a substantial share of the sample (67%), while most others are comparatively small: the energy crisis is picked up by small regimes that shift quickly from one to another. This uneven coverage matters for interpretation. For example, if we were to make inferences about prices at low residual loads in regime 4, the exercise would be completely an extrapolation, and the resulting estimates would be less reliable.

Despite the large differences in terms of price levels, the main transformation is not a simple vertical shift of the same curve. All regimes include hours with zero or slightly negative prices, so the left-hand starting point is broadly similar, and it is the upper parts of the curves that are vastly different. Namely, the steepness of the slope changes much more than the intercept.

Turning to the shape of the supply curves, on the one hand, most regimes display the familiar convexity at high residual loads, consistent with the idea that scarcity drives up marginal costs sharply. On the other hand, in a few regimes (for instance, 3 and 9 here), the supply curve appears steeper at the beginning than at the end, but this could simply reflect limited observations in the scarcity region rather than a true structural change. Finally, most regimes with two breakpoints share a common pattern: a steeper initial segment, a flatter middle section, followed with a steeper end. This structure is particularly clear in regimes 1 and 2 and resurfaces again in regime 11, which pushes us to check an assumption we made earlier.

In Section 3.1, we explained our assumption about regimes being time-continuous but left open the question of whether a regime can be identical to a previous one. Besides our theoretical reasoning, we believe this is fundamentally an empirical question. To answer this question, we compare each regime's supply curve with those of all others, using the area between curves as a similarity measure. Figure 7 summarizes these pairwise distances.

On one hand, the results in Figure 7 seem to provide evidence that there is some reversion to previous regimes. There are several instances of regimes that are more similar to regimes that are two steps apart rather than the immediate one. For example, regimes 3 and 5 are more similar to each other than both are with 4; the same thing happens with regimes 4-6 and 7-9. Moreover, at the end of the period, there is a significant return to previous patterns. Regime 2 is more similar to regime 11 than it is to regime 1. It would seem then that after the three years of the energy crisis, electricity supply has got back to its previous state, but on the other hand, we also know that the thermal supply stack has changed quite drastically since then, mainly due to nuclear phase-out.

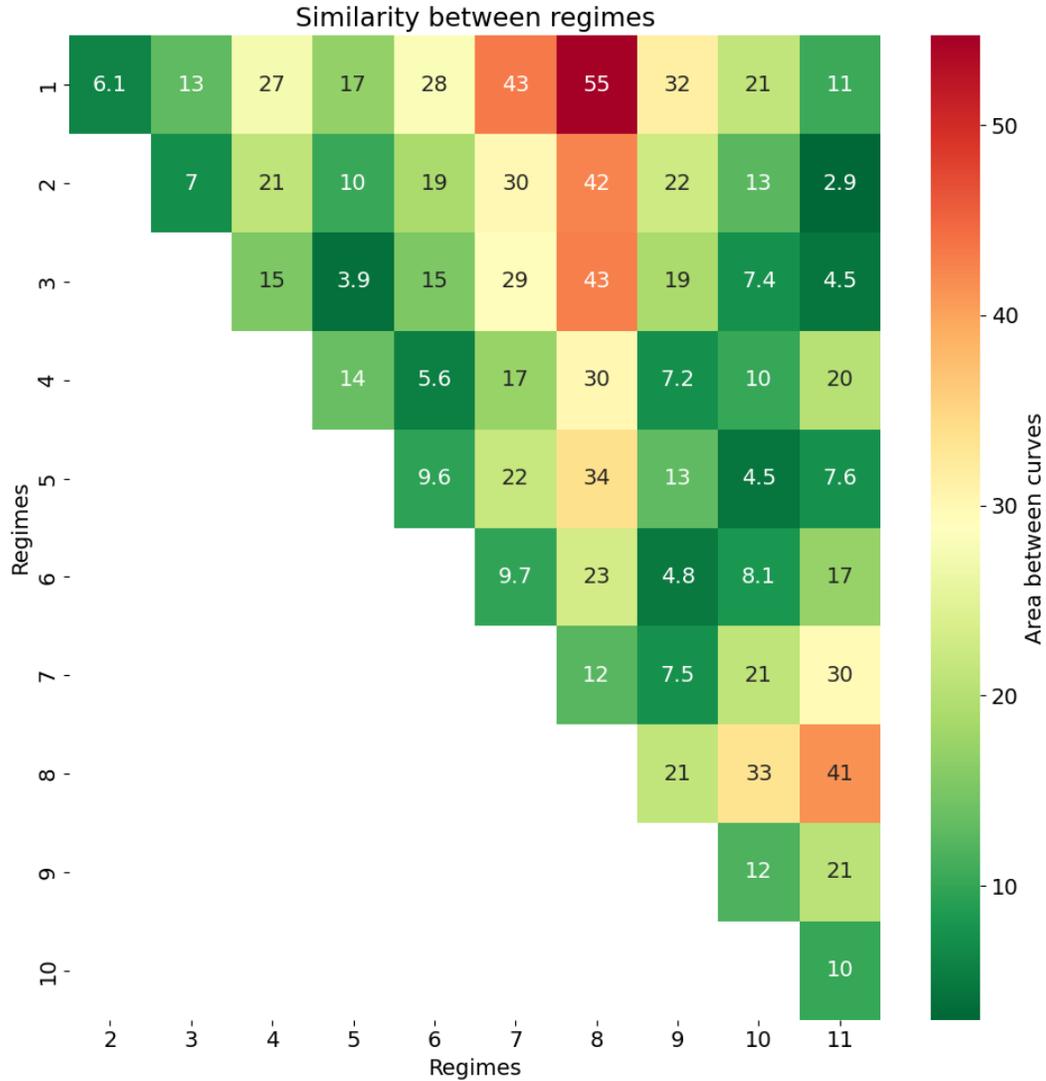

**Figure 7.** Diagonal matrix with pairwise comparisons of the area between supply curves identified by E6. The area between curves is a measure of dissimilarity, where higher numbers (in red) indicate that two curves are further apart, whereas lower numbers (in green) imply more similarity.

## *4.1. Discussion*

The results showcase three key contributions of our approach. First, we show that a simple approach with minimum assumptions can meaningfully identify structural regimes in wholesale electricity supply. The method does not rely on behavioral or engineering modelling; instead, it uses the minimal structure of a monotonic piecewise function to recover empirical supply curves directly from observed equilibria. Second, by comparing cause-driven and effect-driven specifications, we can gauge the relative importance of the underlying cost drivers (causes) in shaping these regimes (effects). The strong alignment between C1 and all effect-driven methods indicates that fuel prices were the dominant drivers of regime shifts in Germany. Third, our study is, to our knowledge, the first to apply a dynamic empirical supply-curve approach to the period of the European energy crisis and the phase-out of nuclear power plant. This provides new evidence on how input-price shocks translate into structural changes in market fundamentals, and highlights that a planned phase-out is not necessarily disruptive to supply.

Our method should be understood for what it is: a tool for generating ex-post empirical supply curves suitable for causal inference and counterfactual analyses. It complements, rather than replace, other approaches in the literature of electricity markets. Studies using bid-level data can better analyze the behavior of market actors, such as strategic behavior, factors driving bidding incentives, and firm-level responses. Engineering dispatch models can study supply stack changes, introduction of new technologies and policy reforms. They are the natural choice for ex-ante analysis of market design or system transformation. Likewise, reduced-form econometric approaches are often more appropriate for isolating the effect of a specific driver (e.g., renewables availability, interconnector flows) on price or quantity outcomes within a sample period.

We argue that our outcome curves are better suited to answer questions about counterfactual market outcomes (price and quantity) ex-post. Moreover, by checking the alignment across specifications, we can answer questions such as whether fuel prices were the dominant driver of regime shifts (yes) or whether cross-border trade significantly affected domestic supply regimes (no). The resulting slope and elasticity estimates can then serve as inputs for general-purpose economic frameworks, such as welfare calculations or pass-through estimates.

We can further learn from comparing cause-driven and effect-driven specifications. Theoretically, the two approaches can diverge in situations where price formation deviates from the observed fundamental drivers. First, it could be that the effect-driven was sensitive to detect a new regime where the cause-driven would not. Policy changes, such as price caps (i.e., the Iberian exception), changes to interconnector capacity, and collusion or anticompetitive behavior are a few examples of factors that could significantly impact price formation. Yet, the cause-driven approach would not pick these up, as the national supply curve fundamentals would not have changed. The reverse situation is also possible, where cause-driven may detect a regime shift that is unobserved in market outcomes. Ultimately, the effect-driven algorithm is unaware of the supply curve outside the region where price formation happens. It may thus occur that a strong increase in gas prices triggers a new cause-driven regime, but – in a country and period where gas power plants are never marginal – does not translate into a new effect-driven one.

It is also worth noting that not all regime changes can be detected by our method. First, supply curves are simplified tools that do not capture temporal interdependencies. But most importantly, our method can only detect phenomena that impact the market outcomes. In the European context there are several issues that are addressed outside the market, such as zonal redispatch. For example, a capacity expansion project within a price zone would not alter the zonal supply curve and hence cannot be detected by looking at market outcomes, even if it does alter the physically available supply.

Lastly, a key strength of the cause-driven method is computational efficiency. It runs quickly, requires few inputs, and seems to capture all major regime changes with only minor timing discrepancies (typically a matter of days). The effect-driven method, in contrast, is more computationally intensive and risks overfitting, since it uses price–quantity equilibria directly to infer the supply curve. But the benefit is that it provides the curves with a better fit to the sets of market equilibria, which is necessary to ensure the exogeneity of our estimate.

With that in mind, we recommend a tailored application of our methods. Cause-driven segmentation is well suited for identifying and interpreting the regime shifts caused by specific economic drivers, while effect-driven segmentation is better suited for recovering the empirical supply curve itself. If both approaches are applied and the results are closely aligned, this should increase confidence that the identified regimes are meaningful. On the contrary, if they diverge, this divergence could itself be informative: it signals the presence of unobserved shocks or drivers affecting price formation independently of observed fundamentals.

## 5. Conclusion

Existing approaches that study wholesale electricity supply –numerical engineering models, bid-based analyses, and parametric econometrics– offer valuable insights but do not provide empirical supply-curve estimates suited for causal inference on market outcomes or welfare analysis, analogous to the empirical studies on electricity demand. This paper contributes a data-driven method to fill that gap.

Building on the work by Tiedemann (2025), we develop an identification strategy based on the concept of stable supply regimes to identify supply curves from market equilibria without using an instrumental variable. We adapt a time partition algorithm developed by Killick et al. (2012) that runs in linear time. We approached the task of time segmentation from two different angles: we looked for stable fundamental drivers of supply (which we argue *should cause* stable supply curves) and for stable price-quantity relationships (which we argue *should be the effect of* stable supply curves). The fact that the identified supply regimes from various specifications concur supports our application of the time partition algorithm. We argue that any researcher interested in producing a supply curve suited for causal inference can do so by adapting our method to their context of interest.

Applying the method to the case of Germany between 2019–2024, we show that supply conditions are often stable for long periods but shifted rapidly during the 2021-2023 energy crises. Our analysis shows that carbon-adjusted coal and gas prices were the main drivers of regime changes in the period, whereas thermal plant availability plays a limited role. The complete nuclear phase-out that took place during the same time does not appear to have destabilized supply. Other potential drivers may introduce short-run noise in price formation but do not appear to define the mid-run supply.

## CRediT authorship contribution statement

Jorge Sánchez Canales: Writing – Original draft, Methodology, Visualization, Software, Project administration, Formal analysis, Conceptualization.

Alice Lixuan Xu: Writing – Original draft, Methodology, Visualization, Conceptualization.

Chiara Fusar Bassini: Writing – Original draft, Methodology, Visualization, Software, Conceptualization.

Lynn H. Kaack: Writing – Review & editing, Funding acquisition, Conceptualization.

Lion Hirth: Writing – Review & editing, Funding acquisition, Conceptualization.

## Data availability

The code and data needed to replicate the analysis are available under an open license and can be found at https://github.com/jscanales/supply_clustering.

## Acknowledgements

This work is supported by the German Federal Ministry of Research, Technology and Space (BMFTR) via the Kopernikus project ARIADNE II (FKZ 03SFK5K0-2) and the ML-Strom Project (FKZ 16DKWN102) and is part of the German Recovery and Resilience Plan (DARP), financed by NextGenerationEU, the European Union's Recovery and Resilience Facility (ARF).

# 7. Appendix A. Pseudocode

---

**Pruned Exact Linear Time segmentation**

**Input**: Sequence of vectors $(\mathbf{y}_1, \mathbf{y}_2, \ldots, \mathbf{y}_n)$, where $\mathbf{y}_i \in \mathbb{R}^m$; cost function $C(.)$; penalty $\beta$.

**Initialize**: $F(0) = -\beta$, $cp(0) = None$, $R_1 = \{0\}$

**Iterate**: For $\tau^* \in \{1, \ldots, n\}$:

Solve: $F(\tau^*) = \min_{\tau \in R(\tau^*)} [F(\tau) + C(\mathbf{y}_{(\tau+1):\tau^*}) + \beta]$.

Set: $\tau^1 = \arg\{\min_{\tau \in R(\tau^*)} [F(\tau) + C(\mathbf{y}_{(\tau+1):\tau^*}) + \beta]\}$,

$cp(\tau^*) = [cp(\tau^1), \tau^1]$,

$R_{(\tau^*+1)} = \{\tau \in R_{\tau^*} \cup \{\tau^*\}: F(\tau) + C(\mathbf{y}_{(\tau+1):\tau^*}) \leq F(\tau^*)$

**Output**: $cp(n)$, final set of changepoints.

---

**Algorithm 1.** PELT from Killick et al. (2012). Original implementation includes an additional constant $K$, which depends on the cost function and controls the strictness of pruning. As the deployed cost functions are based on the sum of squared errors, $K$ can be set equal to zero in this application.

---

**Non-negative Piecewise Linear Fit**

**Input**: Vectors $\mathbf{x}, \mathbf{y} \in \mathbb{R}^n$; number of breakpoints $k$

**Define**: $A \in \mathbb{R}^{n \times k}$ with $A\{i,j\} := \begin{cases} 1 & \text{if } j = 1 \\ (x_i - b_{j-1}) \mathbb{1}_{x_i > b_{j-1}} & \text{if } j \neq 1 \end{cases}$

$\mathbf{b} \in \mathbb{R}^k$ a candidate vector of breakpoints such that:

$$\min\{\mathbf{x}\} = b_1 \leq b_2 \leq \cdots \leq b_k = \max\{\mathbf{x}\} \quad (1)$$

$$\sum_{j'=2}^{j} b_{j'} \geq 0 \qquad \forall j \in \{2, \ldots, k\} \quad (2)$$

$SSR(\mathbf{b}) = (A\mathbf{b} - \mathbf{y})^T (A\mathbf{b} - \mathbf{y})$, the sum of squared errors for a constrained least square optimization.

**Output**: $\hat{f}(x)$ the optimal continuous piecewise linear function

---

**Algorithm 2.** Non-negative Piecewise Linear Fit, adapted from the implementation of PWLF from Jekel and Venter (2019). Our algorithm builds on the original implementation for a fixed number of breakpoints to retrieve curves with non-negative slopes. This condition is enforced through constraint (2).

# 8. Appendix B. Additional Figures

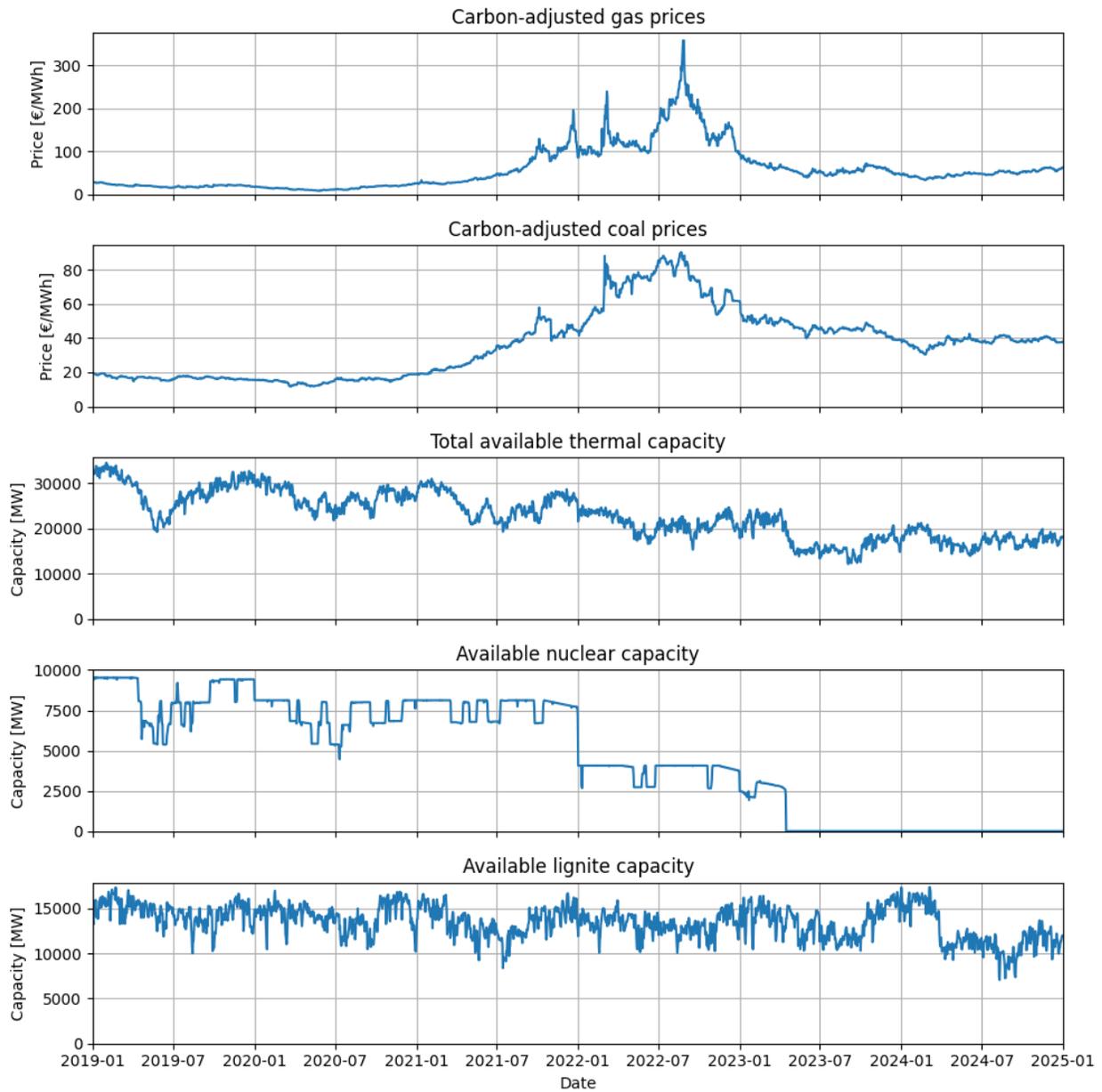

**Figure 8.** Shows the five input series we use in cause-driven specifications in their respective levels. Gas and coal prices show the high spikes that constituted the European energy crises of 2021-2023. Total thermal capacity shows a strong seasonal component, the result of environmental regulations and companies scheduling yearly maintenance during periods of low residual load. At the same time there is a decreasing trend. A sizeable part of this corresponds to the nuclear phase out, that can be seen in the fourth panel. Meanwhile, lignite varies more during the year and across days, but there isn't such a significant long-term trend.

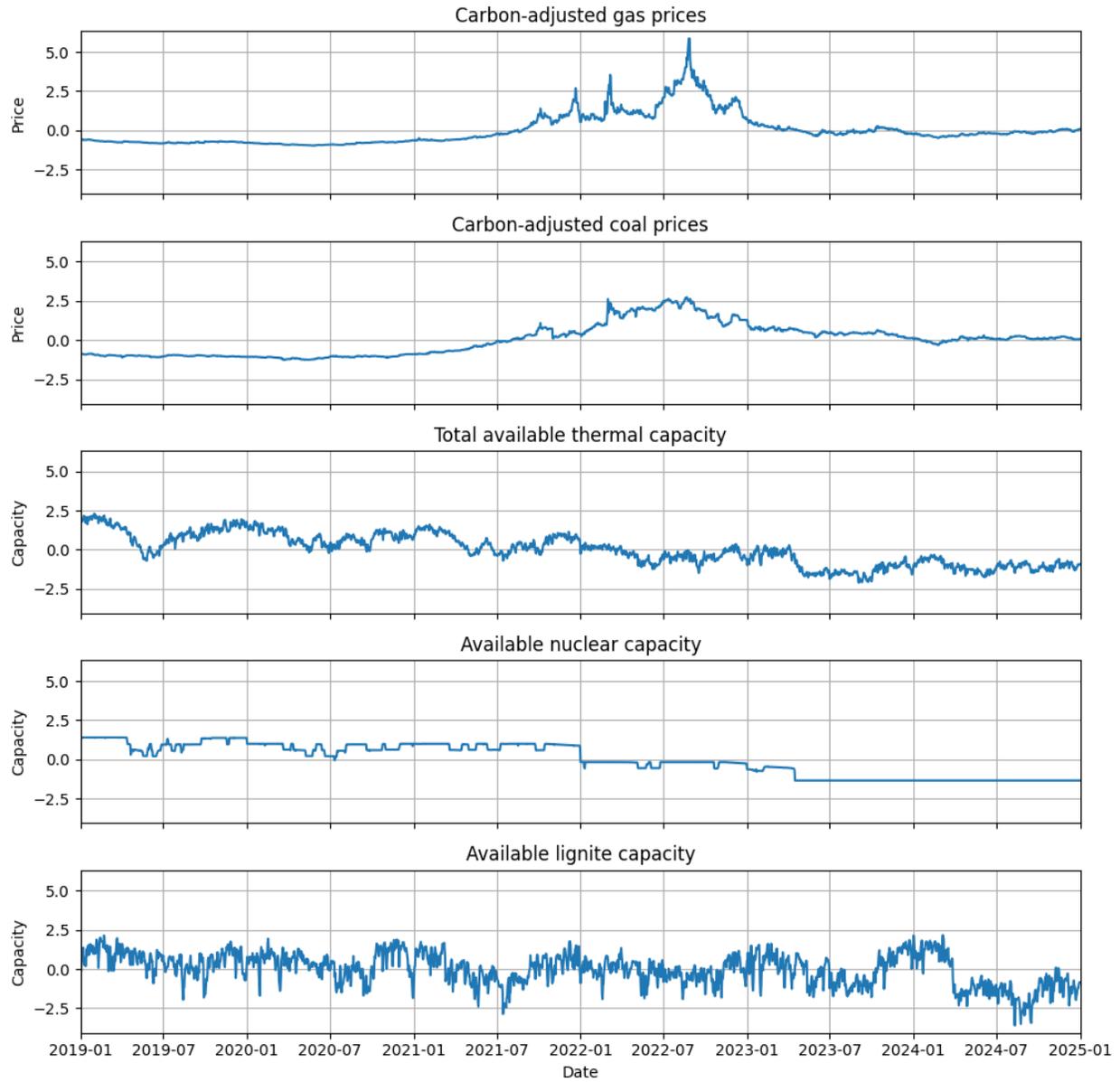

**Figure 9.** Shows the five input series we use in cause-driven specifications. The data here has been normalized, so that our cost function does not outweigh any series. The normal transformation was done by subtracting the mean and dividing by the standard deviation of the series.

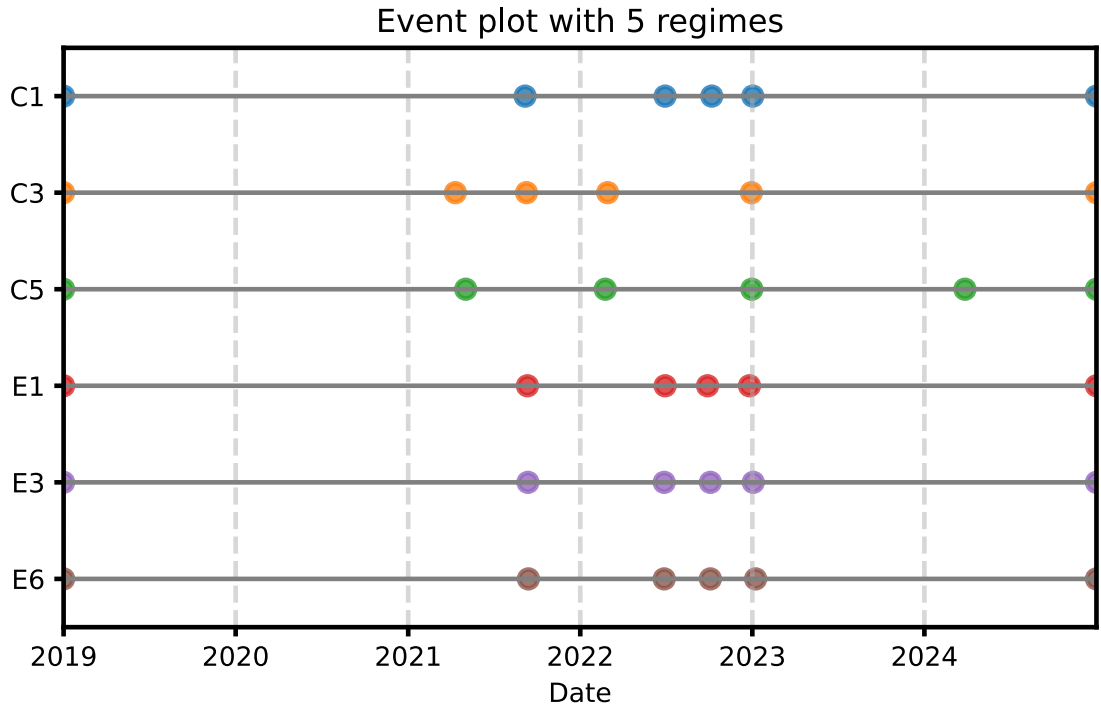

**Figure 10.** Event plot of the whole sample (2019-2024) showing the point in time where each specification identified the regime shifts when looking for 5 regimes.

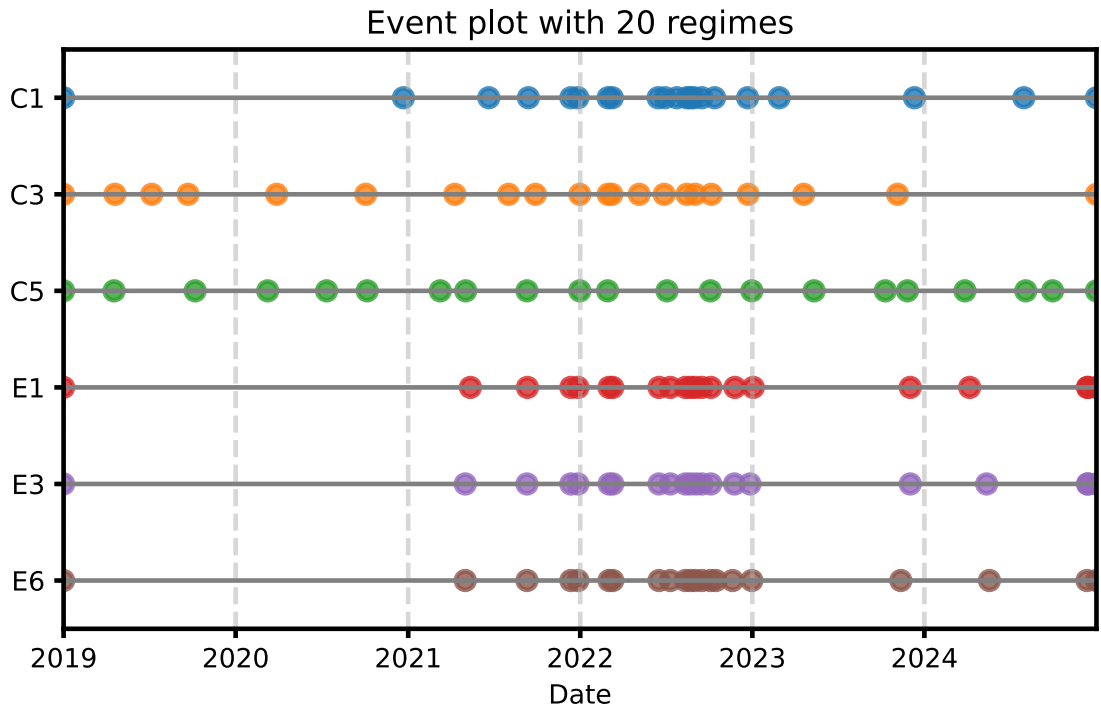

**Figure 11.** Event plot of the whole sample (2019-2024) showing the point in time where each specification identified the regime shifts when looking for 20 regimes.

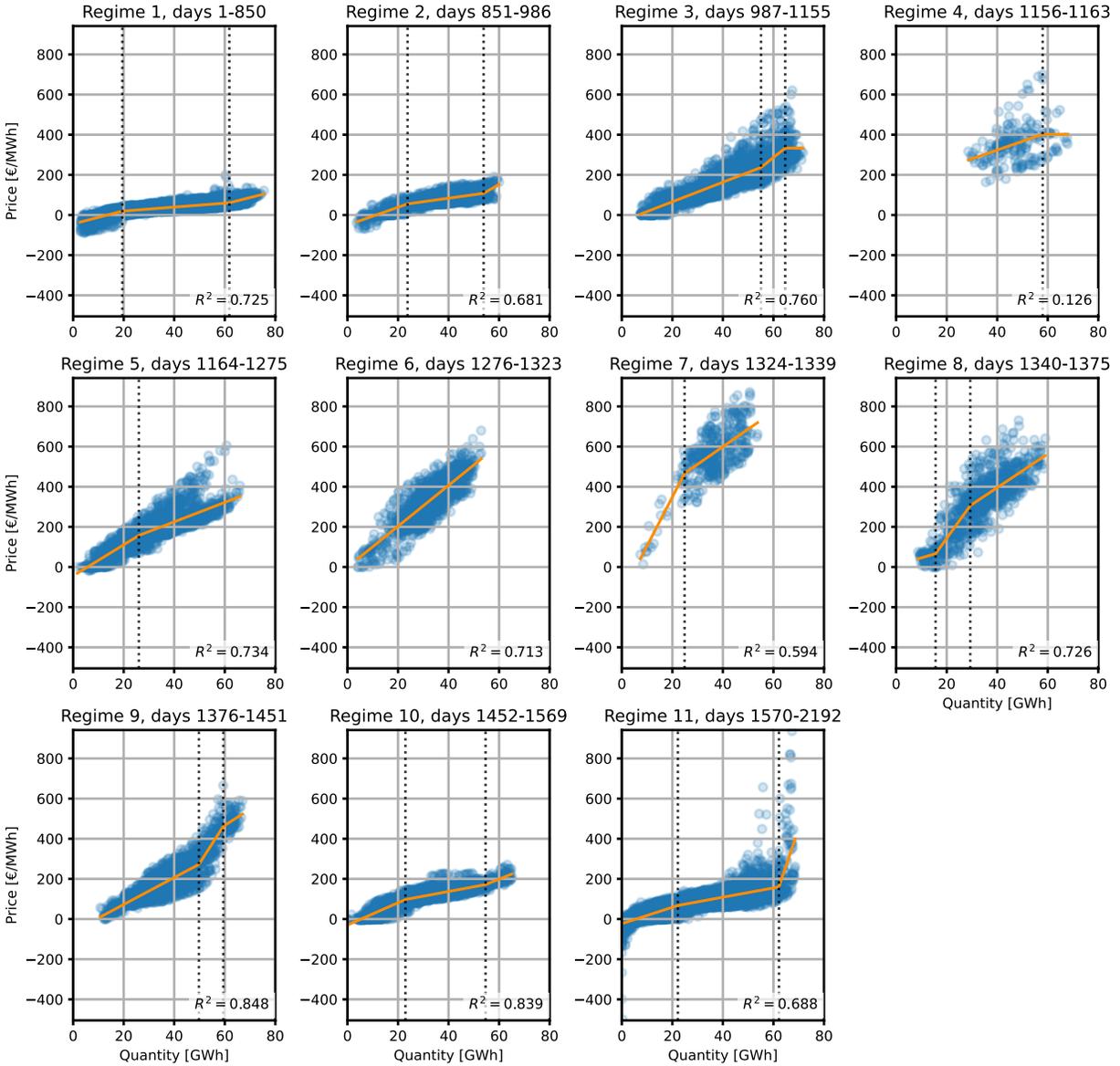

**Figure 12**. Piecewise linear supply curves resulting from using 11 regimes identified by C1. Blue dots represent each hourly equilibrium of the day-ahead market auction. In orange, the estimated piecewise linear supply curve. Dotted vertical lines mark the breakpoints of the piecewise linear functions.

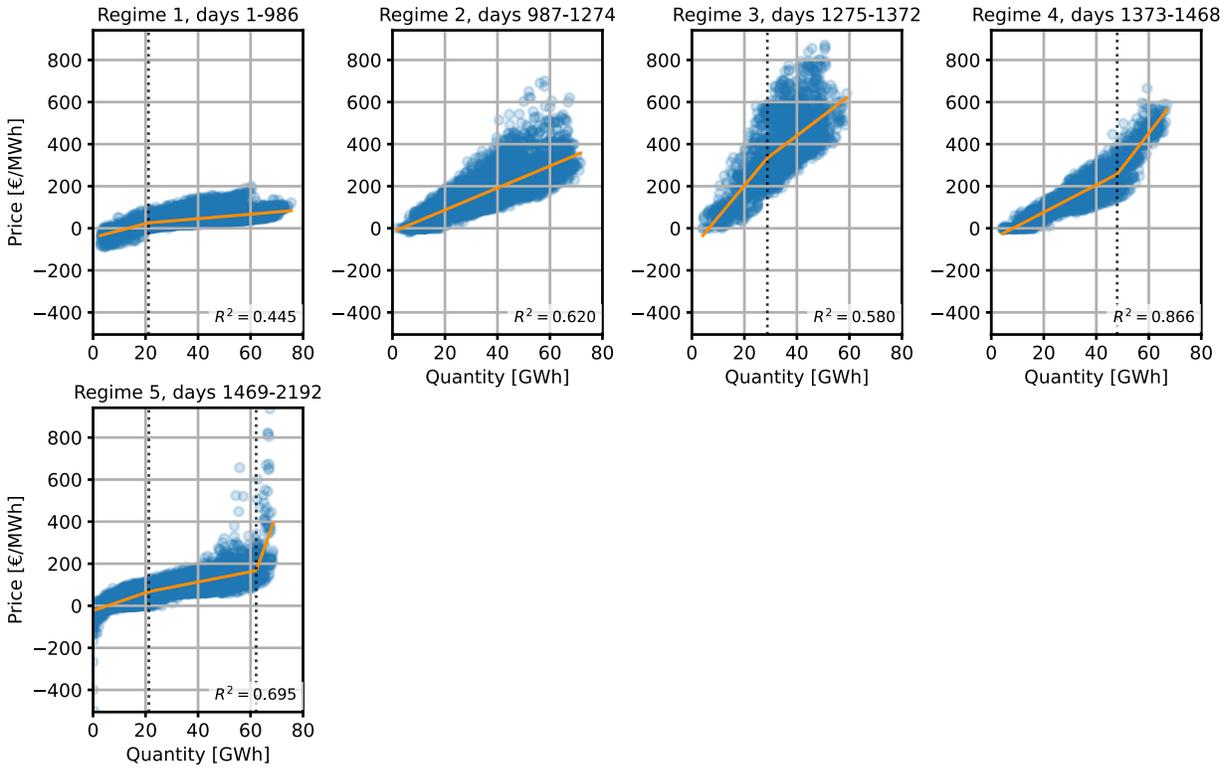

**Figure 13**. Piecewise linear supply curves resulting from using 5 regimes identified by E3. Blue dots represent each hourly equilibrium of the day-ahead market auction. In orange, the estimated piecewise linear supply curve. Dotted vertical lines mark the breakpoints of the piecewise linear functions.

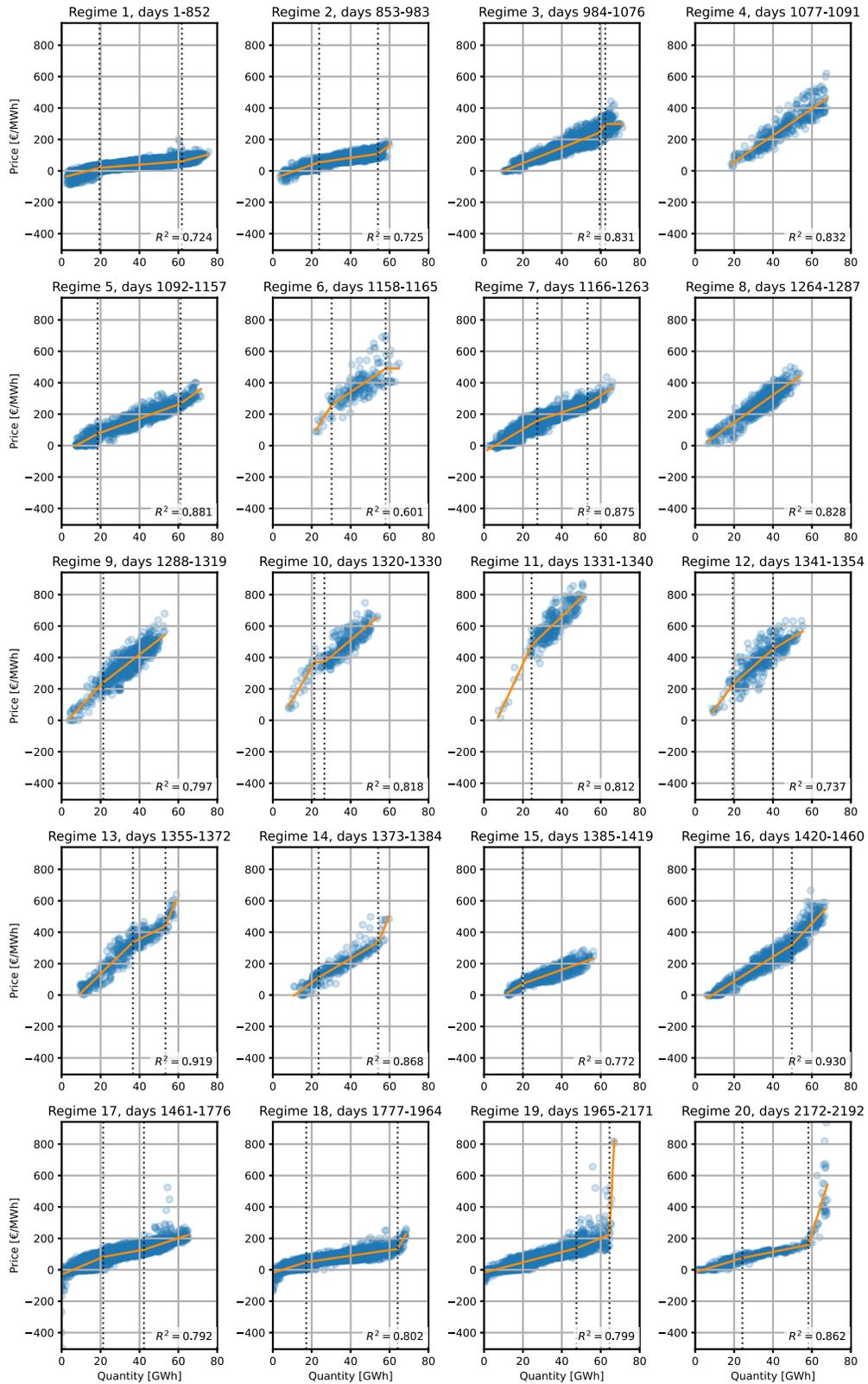

**Figure 14**. Piecewise linear supply curves resulting from using 20 regimes identified by E3. Blue dots represent each hourly equilibrium of the day-ahead market auction. In orange, the estimated piecewise linear supply curve. Dotted vertical lines mark the breakpoints of the piecewise linear functions.